\documentclass[prl,aps,twocolumn,floatfix,showpacs]{revtex4}
\usepackage{graphicx,graphics,psfrag,amsmath,calc}
\usepackage{epsfig}
\usepackage{color}
\begin{document}
\title{
Ultra-cold fermions in the flatland:
evolution from BCS to Bose \\ superfluidity
in two-dimensions with spin-orbit and Zeeman fields}

\author{Li Han and C. A. R. S{\'a} de Melo}
\affiliation{School of Physics, Georgia Institute of Technology,
Atlanta, Georgia 30332, USA}
\date{\today}

\begin{abstract}
We discuss the evolution from BCS to Bose superfluidity for ultracold
fermions in two-dimensions and in the presence of simultaneous
spin-orbit and Zeeman fields. We analyze several thermodynamic
properties to characterize different superfluid phases including pressure,
compressibility, induced polarization, and spin susceptibility.
Furthermore, we compute the momentum distribution and construct
topological invariants for each of the superfluid phases.
\end{abstract}
\maketitle

%
%

Ultra-cold Fermi atoms are one of the most interesting physical
systems of the last decade, as they have served as quantum simulators
of crossover phenomena and phase transitions
encountered in several areas of physics.
Due to their tunable interactions,
atoms like $^6$Li and $^{40}$K have been used to study
the crossover from BCS to BEC superfluidity,
to simulate superfluidity in neutron stars,
and to investigate unitary interactions which are of great
interest in nuclear physics.
%
%
Furthermore, the ability to control the internal spin state of
the atoms by using
radio-frequencies (RF) enabled the studies of quantum and classical phase
transitions as a function of interactions and population imbalance.
%
%
These tools have permitted the study of crossover phenomena and
phase transitions, and have validated the symmetry based
classification of phase transitions put forth by Landau over the
thermodynamic classification proposed earlier by Ehrenfest.

%
%

Very recently three new tools have been developed for ultra-cold fermions.
The first tool is a method that allows for the extraction of
thermodynamic properties - such as
pressure, entropy, compressibility and spin-susceptibility -
for uniform systems by using local density
profiles~\cite{salomon-2011, zwierlein-2011}.
The second tool provides the ability
to trap two dimensional fermions in the quantum degenerate
regime~\cite{turlapov-2010, kohl-2011}.
The third tool is the creation of artificial spin-orbit and Zeeman fields~\cite{spielman-2011},
which allows for the visitation of
previously inaccessible physical regimes,
such as the limit of strong spin-orbit coupling (SOC).
Thus far, the SOC created is of the equal
Rashba-Dresselhaus (ERD) type
instead of Rashba-only~\cite{rashba-1984}
or Dresselhaus-only~\cite{dresselhaus-1955}.
The emergence of the third tool in the context of ultra-cold
bosons~\cite{spielman-2011} has generated substantial theoretical
interest about the possibility of applying the technique
to three-dimensional ultra-cold
fermions~\cite{chapman-sademelo-2011, shenoy-2011,
chuanwei-2011, hu-2011, zhai-2011, iskin-2011, han-2012, seo-2012}.
Several groups studied the Rashba SOC~\cite{shenoy-2011, chuanwei-2011, hu-2011, zhai-2011,
iskin-2011}, which occurs naturally in condensed matter physics,
while our group explored the ERD case~\cite{han-2012, seo-2012}.
In the context of ultra-cold fermions, the combination of
the first two new tools has been achieved experimentally very recently~\cite{zwierlein-2012a},
and there are recent reports that ERD SOC
has been created in fermionic systems of $^{40}$K~\cite{chinese-2012} and $^{6}$Li~\cite{zwierlein-2012b}.

In this manuscript, we discuss the combination of these three new
techniques to describe the evolution from BCS to Bose superfluidity
in two-dimensions, which in the context of ultra-cold fermions has been
discussed for population balanced $s$-wave~\cite{botelho-2006},
$p$-wave~\cite{botelho-2005a}, and imbalanced $s$-wave~\cite{tempere-2009}
systems without SOC. Preliminary accounts of
the effects of Rashba SOC have recently appeared in the
cold atom literature~\cite{wei-2011, chuanwei-2011}.
%
%
Here, we study the simultaneous effects of ERD spin-orbit and
Zeeman fields, which produce a very rich phase diagram consisting of
different topological superfluid. To describe
these novel phases, we discuss their spectroscopic and
thermodynamic properties.

%
%

%
%
%

%
%

%
%

%
%

%
%

%
%

{\it Hamiltonian:}
We start with the total Hamiltonian
$
H
=
H_{\rm sp}
+
H_{\rm int},
$
where the single-particle part is ($\hbar = 1$)
\begin{equation}
\label{eqn:hamiltonian-single-particle}
H_{\rm sp} =
\sum_{{\bf k}, s, s^\prime, i}
\psi^{\dagger}_{s} ({\bf k})
\left[
 \hat{K} \delta_{s s^\prime}
- h_i ({\bf k}) \sigma_{i, s s^\prime}
\right]
\psi_{s^\prime} ({\bf k}).
\end{equation}
Here, ${\hat K} = k^2/(2 m) - \mu$
is the kinetic energy in reference to the chemical potential
$\mu$ and $h_i ({\bf k})$ is the spin-orbit plus Zeeman field along
the $i$-direction ($s, s^\prime = \uparrow, \downarrow$, $i = x, y, z$).
In the present case, we consider only the
ERD SOC field
${\bf h}_{\rm ERD} = (0, v k_x, 0)$~\cite{spielman-2011, han-2012, seo-2012}
and Zeeman field ${\bf h}_{\rm ZEE} = (0, h_y, h_z)$. The current
Hamiltonian matrix that can be created in the laboratory
using Raman beams is of the form
$
H_{\rm ZSO}
=
- h_z\sigma_z
- h_y  \sigma_y
- h_{\rm ERD}\sigma_y,
$
where $h_z = (\Omega/2)$ with $\Omega$ being
the Raman intensity, $h_y = (\delta/2)$ with $\delta$ being
the detuning, and $h_{\rm ERD} = v k_x$ with $v$ being a measure of the
SOC strength.

The interaction Hamiltonian is
\begin{equation}
H_{\rm int}
=
\sum_{{\bf k}, {\bf k}^\prime, {\bf q}}
V_{{\bf k}{\bf k}^\prime}
b_{\bf q}^\dagger ({\bf k})
b_{\bf q} ({\bf k}^\prime),
\end{equation}
where
$
b_{\bf q}^\dagger ({\bf k})
=
\psi^{\dagger}_{\uparrow} ({\bf k} + {\bf q}/2)
\psi^{\dagger}_{\downarrow} (-{\bf k} + {\bf q}/2)
$
is the creation operator for a pair of fermions with center of mass
momentum ${\bf q}$ and relative momentum $2{\bf k}$.
The interaction term $V_{{\bf k} {\bf k}^\prime}$ is
the Fourier transform of the contact interaction
$V ({\bf r}, {\bf r}^\prime)
=
-
g \,
\delta ({\bf r} - {\bf r}^\prime)
$ ($g > 0$).

%
%
%
%

The single-particle Hamiltonian $H_{\rm sp}$ can be diagonalized
by a momentum dependent SU(2) transformation
into the helicity basis
$
\Phi^\dagger ({\bf k})
=
M ({\bf k}) \Psi^\dagger ({\bf k})
$,
where the helicity creation operators
$
\Phi^\dagger ({\bf k})
=
\left(
\phi_{\Uparrow}^\dagger ({\bf k}),
\phi_{\Downarrow}^\dagger ({\bf k})
\right)
$
are expressed as linear combinations of the
standard creation operators
$
\Psi^\dagger ({\bf k})
=
\left(
\psi_{\uparrow}^\dagger ({\bf k}),
\psi_{\downarrow}^\dagger ({\bf k})
\right).
$
The eigenvalues of the Hamiltonian matrix ${\bf H}_{\rm sp} ({\bf k})$ are
%
$
\xi_{\Uparrow, \Downarrow} ({\bf k})
=
K ({\bf k})
\mp
\vert {\bf h}_{\rm eff} ({\bf k})\vert,
$
%
corresponding to the helicity spin $\Uparrow$ ($\Downarrow$)
being aligned (anti-aligned) with respect to the effective magnetic field
$
{\bf h}_{\rm eff} ({\bf k})
= (0, h_{\rm ERD} ({\bf k}), h_z)
$.
Here, the magnitude of the effective field is
$
\vert {\bf h}_{\rm eff}({\bf k})\vert
=
\sqrt
{
h_z^2
+
v^2 k_x^2
}
$
and
$
K ({\bf k}) = \epsilon_{\bf k} - \mu
=
k^2
/
(2m)
-
\mu
$.

In the helicity basis, the interaction Hamiltonian transforms as
%
$
H_{\rm int}
=
-g
\sum_{{\bf q} \alpha \beta \gamma \delta}
B^\dagger_{\alpha \beta} ({\bf q})
B_{\gamma \delta} ({\bf q}),
$
%
where $\alpha, \beta, \gamma, \delta = \Uparrow, \Downarrow$. Pairing is now described
by the operator
$
B_{\alpha \beta}^\dagger ({\bf q})
=
\sum_{\bf k}
\Lambda_{\alpha \beta}
({\bf k}_1, {\bf k}_2)
\Phi_{\alpha}^\dagger ({\bf k}_1 )
\Phi_{\beta}^\dagger  ({\bf k}_2)
$
and its Hermitian conjugate, with momentum indices
${\bf k}_1 = {\bf k} + {\bf q}/2$
and
${\bf k}_2 = -{\bf k} + {\bf q}/2$.
Pairing occurs between fermions of momenta
${\bf k}_1$ and ${\bf k}_2$ in two different helicity
bands (inter-helicity pairing), as well as within the same helicity
band (intra-helicity pairing).
The tensor
$
\Lambda_{\alpha \beta}
({\bf k_1}, {\bf k_2})
$
contains the matrix elements of the momentum
dependent SU(2) rotation
into the helicity basis,
and reveals that the center of mass momentum
${\bf k}_1 + {\bf k}_2 = {\bf q}$ and
the relative momentum ${\bf k}_1 - {\bf k}_2 = 2{\bf k}$
are coupled and no longer independent.

%
%

%
%
For pairing at ${\bf q} = 0$,
the order parameter for superfluidity is the tensor
$
\Delta_{\alpha \beta} ({\bf k})
=
\Delta_0
\Lambda_{\alpha \beta} ({\bf k}, -{\bf k}),
$
where
$
\Delta_0
=
- g
\sum_{\gamma \delta}
\langle
B_{\gamma \delta} ({\bf 0})
\rangle,
$
leading to
components:
$
\Delta_{\Uparrow \Uparrow} ({\bf k})
=
i \widetilde\Delta_T ({\bf k})
$
for helicity projection
$\lambda = +1$;
$
\Delta_{\Uparrow \Downarrow} ({\bf k})
= - \Delta_{\Downarrow \Uparrow} ({\bf k})
= -
\widetilde\Delta_S ({\bf k})
$
for helicity projection $\lambda = 0$; and
$
\Delta_{\Downarrow \Downarrow} ({\bf k})
=
- i \widetilde\Delta_T ({\bf k})
$
for helicity projection $\lambda = -1$.
The amplitudes
$
\widetilde\Delta_T ({\bf k})
=
\Delta_0 h_{ERD} ({\bf k})
/
\vert {\bf h}_{\rm eff} ({\bf k}) \vert
$
and
$
\widetilde\Delta_S ({\bf k})
=
\Delta_0
h_z
/
\vert {\bf h}_{\rm eff} ({\bf k}) \vert
$
reflect the triplet and singlet components
of the order parameter in the helicity basis.
For the ERD case, $\widetilde \Delta_T ({\bf k})$ is odd and
$\widetilde \Delta_S ({\bf k})$ is even under parity ${\cal P}$ in momentum space,
while they are both invariant under time reversal
${\cal T}$, and both reflect the broken U(1)
symmetry of the superfluid phase.
The triplet and singlet sectors in the
helicity basis are not independent as the
interactions from which they originate occur only
in the singlet $s$-wave of the original spin basis, and the relation
$
\vert \widetilde\Delta_T ({\bf k}) \vert^2
+
\vert \widetilde \Delta_S ({\bf k}) \vert^2
=
\vert \Delta_0 \vert^2
$
holds.

%
%
%
%

It is worth emphasizing that in the triplet sector
$\Delta_{\Uparrow \Uparrow} ({\bf k}) $ and
$\Delta_{\Downarrow \Downarrow} ({\bf k})$
contain not only $p$-wave, but also $f$-wave and higher
odd partial waves in two dimensions,
as can be seen from a {\it multipole}
expansion of
$
\vert {\bf h}_{\rm eff} ({\bf k}) \vert^{-1}
=
\left[
h_z^2 + h_{\rm ERD}^2 ({\bf k})
\right]^{-1/2}
$
for finite $h_z$.
Similarly in the singlet sector
$\Delta_{\Uparrow \Downarrow} ({\bf k})$ and
$\Delta_{\Downarrow \Uparrow} ({\bf k})$
contain $s$-wave, $d$-wave and
higher even partial waves, as long
as the Zeeman field $h_z$ is non-zero.
Higher angular momentum pairing
occurs because the local (zero-ranged) interaction in the
$(\uparrow, \downarrow)$
spin basis is transformed into a finite-ranged anisotropic
interaction in the helicity basis $(\Uparrow, \Downarrow)$.

%
%

%
%

The effective Hamiltonian in the helicity basis
takes the matrix form
\begin{equation}
\label{eqn:saddle-point-hamiltonian-helicity-basis}
H_{\rm eff} ({\bf k})
=
\left(
\begin{array}{cccc}
\xi_{\Uparrow}({\bf k}) & 0 &
\Delta_{\Uparrow \Uparrow} ({\bf k}) & \Delta_{\Uparrow \Downarrow} ({\bf k}) \\
0  & \xi_{\Downarrow}({\bf k})&
\Delta_{\Downarrow \Uparrow} ({\bf k}) & \Delta_{\Downarrow \Downarrow} ({\bf k}) \\
\Delta_{\Uparrow \Uparrow}^* ({\bf k}) & \Delta_{\Downarrow \Uparrow}^* ({\bf k})&
- \xi_{\Uparrow}({\bf k}) &  0 \\
\Delta_{\Uparrow \Downarrow}^* ({\bf k}) & \Delta_{\Downarrow\Downarrow}^* ({\bf k}) &
0  & -\xi_{\Downarrow}({\bf k})
\end{array}
\right).
\end{equation}
The eigenvalues for the quasiparticle bands are
\begin{equation}
E_{p \pm} ({\bf k})
=
\sqrt{
\left(
E_S ({\bf k})
\pm
\vert
{\bf h}_{\rm eff} ( {\bf k} )
\vert
\right)^2
+
\vert \Delta_T ({\bf k}) \vert^2
},
\end{equation}
where
$
E_S ({\bf k})
=
\sqrt{
\vert
K({\bf k})
\vert^2
+
\vert \widetilde\Delta_S ({\bf k}) \vert^2
}
$
is a characteristic energy for the singlet sector.
The eigenvalues for quasihole bands are
$
E_{h\pm} ({\bf k})
=
-
E_{p\pm} ({\bf k}).
$
The effective field $\vert {\bf h}_{\rm eff} ({\bf k}) \vert$ emerges from the energy difference between the helicity energies
$
\vert
{\bf h}_{\rm eff} ({\bf k})
\vert
=
\left[
\xi_\Downarrow ({\bf k})
-
\xi_\Uparrow ({\bf k})
\right]
/
2
$,
while
$
K({\bf k})
=
\left[
\xi_\Downarrow ({\bf k})
+
\xi_\Uparrow ({\bf k})
\right]
/
2
$
is the average energy of the helicity
bands.

Only $E_{p-} ({\bf k})$
has zeros or nodes, provided that the following
conditions are satisfied simultaneously:
first, the effective magnetic field energy
must equal to the {\it excitation energy} for
the singlet sector
$
\vert
{\bf h}_{\rm eff} ({\bf k})
\vert
=
E_S ({\bf k});
$
second, the triplet component of the
order parameter must vanish
$
\vert
\widetilde \Delta_T ({\bf k})
\vert
=
0.
$
The nodal structure for $E_{p-} ({\bf k})$ dominates the physics
at low energies. For the ERD case,
$\vert h_{\rm ERD} ({\bf k}) \vert = v \vert k_x \vert$. Therefore,
zeros of $E_{p-} ({\bf k})$ must occur when $k_x = 0$, which gives rise to
up to four possible point nodes that are determined by the real solutions of $k_y$ to the equation
$\left( k_y^2/(2m) - \mu \right)^2 + \vert \Delta_0 \vert^2 = h_z^2$.
%

%
%

To obtain the phase diagram,
we calculate the thermodynamic potential
$
\Omega
=
-
T \ln
\left[
{\rm Tr} e^{-H_{\rm eff}/T}
\right],
$
and solve self-consistently the order parameter equation
$\partial \Omega/\partial \vert \Delta_0 \vert^2 = 0$
and the number equation
$
N
=
-
\left(
\partial \Omega/\partial \mu
\right)_{T,V}
$.
The contact interaction $g/V$ is expressed
in terms of the two-body binding energy $E_b$
in the absence of SOC via
the relation $V/g = \sum_{\bf k} 1/(2\epsilon_{\bf k} + E_b)$.
%
%

%
\begin{figure} [tbh]
\centering
\epsfig{file=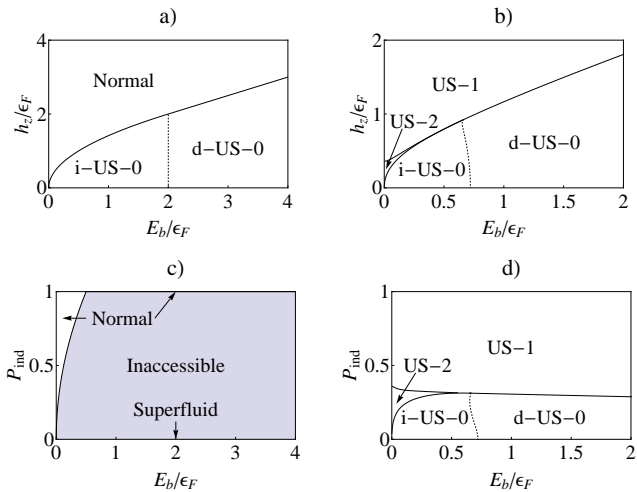,width=1.0 \linewidth}
%
%
\caption{ \label{fig:one}
Zero temperature $(T = 0)$ phase diagrams of Zeeman
field $h_z/\epsilon_F$ versus two-body binding energy
$E_b/\epsilon_F$ for equal Rashba-Dresselhaus (ERD)
SOC (a) $v/v_{F} = 0$ and
(c) $v/v_{F} = 0.8$. The corresponding phase diagrams
for the induced population imbalance $P_{\rm ind}$ versus
$E_b/\epsilon_F$ are shown in (b) $v/v_{F} = 0$ and
(d) $v/v_{F} = 0.8$. Here, $\epsilon_F = k_F^2/(2m)$ and $v_F = k_F/m$ are the Fermi energy and momentum respectively,
and $k_F$ is related to the particle density $n$ via $k_F = (3\pi^2 n)^{1/3}$.
}
\end{figure}

In Fig.~\ref{fig:one}, we show the zero temperature
phase diagrams of Zeeman field
$h_z$ versus the two-body binding energy $E_b$
as well as induced population imbalance $P_{\rm ind}$
versus binding energy for both vanishing and large finite SOC ($v/v_F$ = 0 and 0.8).
We label the uniform superfluid phases with zero,
one or two pairs of nodes as US-0, US-1, and US-2, respectively.
The US-2/US-1 phase boundary is determined
by the condition $\mu = \sqrt{h_z^2 - \vert \Delta_0 \vert^2}$,
when $\vert h_z \vert > \vert \Delta_0 \vert$;
the US-0/US-2 boundary is determined by
the Clogston-like condition $\vert h_z \vert = \vert \Delta_0 \vert$
when $\mu > 0$, where the gapped US-0 phase
disappears leading to the gapless US-2 phase;
and the US-0/US-1 phase boundary is determined by
$\mu = - \sqrt{h_z^2 - \vert \Delta_0 \vert^2}$,
when $\vert h_z \vert > \vert \Delta_0 \vert$.
Furthermore, within the US-0 phase,
a crossover line between an indirectly gapped and
a directly gapped US-0 phase occurs at $\mu = 0$.
Non-uniform (NU) phases also emerge in regions where uniform phases
are thermodynamically unstable for $v/v_F \ne 0$, however NU phases
only exist in the regime of $0 < v/v_F < (v/v_F)_c$, where
$(v/v_F)_c \approx 0.6$.
Possible NU phases include phase separation,
modulated superfluids or supersolid. These possible phases are not illustrated
in Fig.~\ref{fig:one} since we are interested in comparisons between
the zero SOC case $v/v_F = 0$ and the large SOC case
$v/v_F > (v/v_F)_c$. In the latter case, a tri-critical point occurs at the intersection of the US-0, US-1, and US-2
(where $\mu = 0$ and $\vert h_z \vert = \vert \Delta_0 \vert$).

\begin{figure} [tbh]
\centering
\epsfig{file=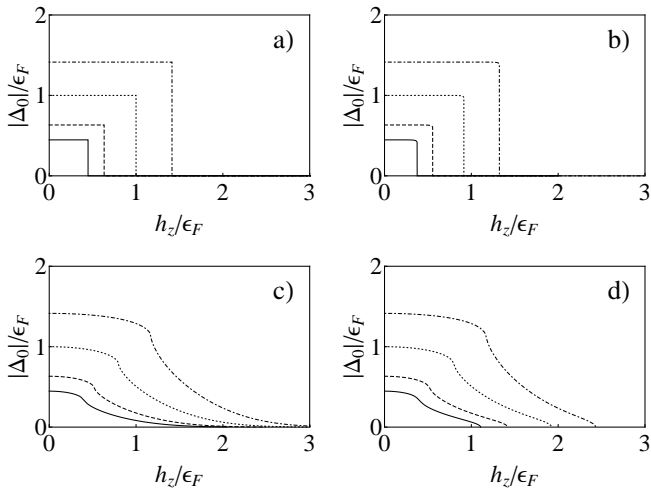,width=1.0 \linewidth}
\caption{
\label{fig:two}
The order parameter amplitude $\vert \Delta_0 \vert/\epsilon_F$
as a function of external Zeeman field $h_z/\epsilon_F$ for
(a) $v/v_F = 0$, $T/\epsilon_F = 0$,
(b) $v/v_F = 0$, $T/\epsilon_F = 0.03$,
(c) $v/v_F = 0.8$, $T/\epsilon_F = 0$,
(d) $v/v_F = 0.8$, $T/\epsilon_F = 0.03$
at various binding energies
$E_b/\epsilon_F = 0.1$ (solid line),
$E_b/\epsilon_F = 0.2$ (dashed line),
$E_b/\epsilon_F = 0.5$ (dotted line),
$E_b/\epsilon_F = 1.0$ (dot-dashed line).
}
\end{figure}

It is important to note from the phase diagrams that
for finite SOC ($v/v_F > 0$) it is always possible to form pairs
in the lower helicity band $\xi_{\Uparrow} ({\bf k})$ no matter
how large the Zeeman field $h_z$ is. This is because an induced triplet
component of the order parameter emerges and circumvents the
standard pair breaking Clogston limit for singlet pairing.
Thus, the stable superfluid phase for large $h_z$ at $T = 0$ is the
US-1 phase as shown in Fig.~{\ref{fig:one} provided that $v/v_F \ne 0$.
This is clearly seen in Fig.~\ref{fig:two} where the order parameter
is shown for the cases of $v/v_F = 0$ and $v/v_F = 0.8$.
In the first case $(v/v_F = 0)$, the Clogston limit at $T =0$ occurs
when
$
h_z/\epsilon_F
\sim
\vert \Delta_0 \vert/\epsilon_F
$
as seen in Fig.~\ref{fig:two}a,
where the order parameter jumps discontinously to zero at a critical
value of $(h_z/\epsilon_F)_c$. The discontinuity occurs also at finite
temperatures (Fig.~\ref{fig:two}b), rendering the transition
from the superfluid to the normal phase discontinuous according
to Landau's classification or first order according to Ehrenfest's.
For $v/v_F = 0.8$, the absence of the Clogston paramagnetic
limit at zero temperature is clearly seen in Fig.~\ref{fig:two}c,
where the order parameter vanishes only asymptotically as
$h_z/\epsilon_F \to \infty$ because of the induced triplet component
in the helicity basis. As the temperature is raised, thermal fluctuations
break pairs, the order parameter vanishes, and the normal state is
reached beyond a critical Zeeman field $(h_z/\epsilon_F)_c$,
which is an increasing  function of the binding energy $E_b/\epsilon_F$
for fixed $v/v_F > (v/v_F)_c$. This also occurs for Rashba-only couplings
in two-dimensions~\cite{wei-2011}, and for ERD couplings in three-dimensions~\cite{seo-2012}.

These superfluid phases can be classified according to the topological properties
of their low-energy excitation spectrum $E_{p-} ({\bf k})$. The nodes of
$E_{p -} ({\bf k})$ can be viewed as vortex singularities in momentum space.
Consider the two-dimensional unit vector
$\hat {\bf m} ({\bf k}) = (m_x, m_y)$, where
$m_x({\bf k}) =
\left[
E_S ({\bf k})
-
\vert {\bf h}_{\rm eff} ({\bf k}) \vert
\right]
/E_{p-} ({\bf k})
$
and $m_y ({\bf k}) = \widetilde \Delta_T ({\bf k})/ E_{p-} ({\bf k})$.
The topological charge is the winding (Chern) number
$
N_{w}
=
(2\pi)^{-1}
\oint d\ell \,
{\hat{\bf z}}
\cdot
{\hat {\bf m}}
\times
d{\hat{\bf m}}/d\ell
$,
where the line integral is over any line in the $k_x$-$k_y$ plane
enclosing a zero of $E_{p -} ({\bf k})$. The topological charges take values $\pm 1$.
In the US-2 (US-1) phase the topological charges follow
the sequence $+-+-$ ($+-$) for the four (two) nodal points as we scan
$k_y$ from negative to postive.

\begin{figure} [tbh]
\centering
\epsfig{file=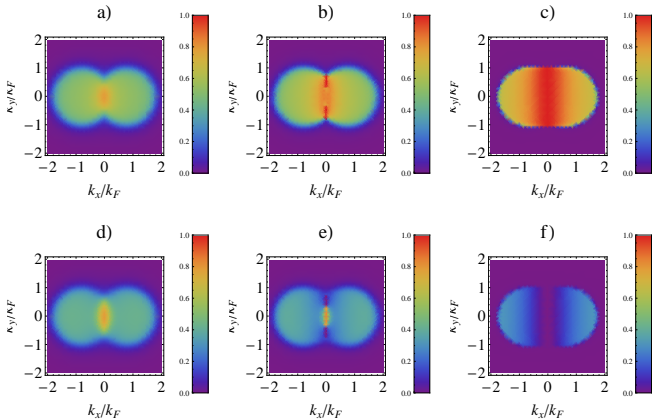,width=1.00 \linewidth}
%
%
\caption{
\label{fig:three}
(color online) The momentum distributions $n_s (k_x, k_y)$
for ERD SOC $v/v_F = 0.8$ and $E_b/\epsilon_F = 0.1$ at $T = 0$,
where $s = \uparrow$ ($\downarrow$) for upper (lower) panels.
(a)(d) i-US-0 phase with $h_z/\epsilon_F = 0.2$;
(b)(e) US-2 phase with $h_z/\epsilon_F = 0.4$;
(c)(f) US-1 phase with $h_z/\epsilon_F = 1.0$.
The color coding varies continuously from purple ($n_s = 0$)
to red ($n_s = 1$).
}
\end{figure}

The change in nodal structure is associated with bulk
topological transitions of the Lifshitz class as noted for
$p$-wave~\cite{volovik-1992, botelho-2005a} and
$d$-wave~\cite{duncan-2000, botelho-2005b} superfluids.
The loss of nodal regions of $E_{p-} ({\bf k})$
correspond to annihilation of Dirac quasi-particles with opposite momenta
and opposite topological charge (US-1/US-2 and US-1/d-US-0),
or with the same momenta but opposite topological charge (US-2/i-US-0).
For instance, starting from the US-1 (US-2) phase,
Dirac quasi-particles annihilate at zero momentum
producing bulk Majorana quasi-particles at the
US-1/d-US-0 (US-2/US-1) phase boundary, which then
become massive Dirac fermions near zero momentum
with positive mass (negative mass) in the d-US-0 (US-1) phase.
However, zero-momentum bulk Majorana quasi-particles do not occur at the
US-2/i-US-0 phase boundary, as the four node superfluid US-2 leads becomes
a doubly-degenerate two-node superfluid at finite momentum
at the phase boundary and then becomes an indirect gap superfluid.

%
%

The emergence of Dirac fermions produces dramatic changes in
momentum distributions~\cite{duncan-2000}
\begin{equation}
n_{{\bf k},s}
=
\frac{1}{2}
\left[
1 -
\sum_{j} n_F [E_j ({\bf k})]
\frac{\partial E_j ({\bf k})}{\partial \mu_s}
\right],
\end{equation}
which are illustrated in Fig.~\ref{fig:three} for parameters $v/v_F = 0.8$ and $E_b/\epsilon_F = 0.1$ at $T =0$.
Different topological phases are shown for different values of Zeeman field $h_z$.
Here, $n_F [E_j ({\bf k})] = 1/(\exp [E_j({\bf k})/T] + 1)$ is the Fermi function.
For the i-US-0 phase the momentum distributions are smooth functions
of ${\bf k} = (k_x, k_y)$ until the i-US-0/US-2 phase boundary is reached where doubly
degenerate nodes emerge at ${\bf k} = (0, \pm \sqrt{2m \mu})$. Beyond this
point discontinuities develop in $n_{{\bf k},s}$, which is due to the finite Zeeman field
and can be best seen in its $\uparrow$ component. Within the US-2
phase there are four discontinuities at the zeros of $E_{p-} ({\bf k})$, which are located at
$
{\bf k} =
\left(0,
\pm
\sqrt{
2m
\left[
\mu \pm \sqrt{h_z^2 - \vert \Delta_0 \vert^2}
\right]
}
\right)
$.
Around these zeros the energy has cusps and the dispersion is linear.
As $h_z$ is increased further, the US-2/US-1 boundary is crossed
and the inner Dirac fermions annihilate, leaving discontinuities in
the momentum distribution occur only at two nodes
$
{\bf k} =
\left(
0,
\pm
\sqrt{
2m
\left[
\mu + \sqrt{h_z^2 - \vert \Delta_0 \vert^2}
\right]
}
\right).
$
The final US-1 phase is highly polarized as can be seen from the great contrast between Fig.~\ref{fig:three}c
and~\ref{fig:three}f.

\begin{figure} [tbh]
\centering
\epsfig{file=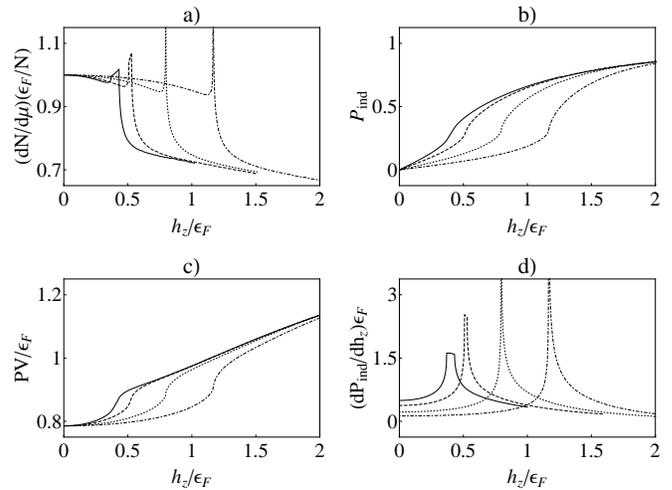,width=1.0 \linewidth}
%
\caption{ \label{fig:four}
Various thermodynamic quantities versus Zeeman field $h_z/\epsilon_F$
for different values of binding energies
$E_b/\epsilon_F = 0.1$ (solid line),
$E_b/\epsilon_F = 0.2$ (dashed line),
$E_b/\epsilon_F = 0.5$ (dotted line),
and $E_b/\epsilon_F = 1.0$ (dot-dashed line).
(a) The pressure $P$; (b) The isothermal compressibility $N^2 \kappa_T$;
(c) The induced polarization $P_{\rm ind}$;
(d) The spin susceptibility $\chi_{zz}$.
}
\end{figure}

We show in Fig.~\ref{fig:four}
several thermodynamic properties as a function of Zeeman field
$h_z/\epsilon_F$ for different binding energies
at $T = 0$.
In Fig.~\ref{fig:four}a, we show the pressure $P = - \Omega/V$.
In Fig.~\ref{fig:four}b, we show the isothermal
compressibility $\kappa_{T} = - V^{-1} (\partial V/\partial P)_T$, which can
also be written as $\kappa_{T} = N^{-2} (\partial N/\partial \mu)_T$.
In Fig.~\ref{fig:four}c and~\ref{fig:four}d, we show the induced
polarization
$
P_{\rm ind}
=
\left(
N_{\uparrow} - N_{\downarrow}
\right)
/
\left(
N_{\uparrow} + N_{\downarrow}
\right)
$,
and the spin susceptibility
$\chi_{zz}
=
\left(
\partial P_{\rm ind}/\partial h_z
\right)_{T,V}.
$
According to Fig.~\ref{fig:one}b, if we fix the value of $E_b$ and increase the Zeeman field $h_z$ from zero,
we may cross several phase boundaries, e.g., from i-US-0 to US-2 to US-1, or from d-US-0 to US-1. Correspondingly,
both $\kappa_T$ and $\chi_{zz}$
are non-analytic at the phase
boundaries (Fig.~\ref{fig:four}c and~\ref{fig:four}d), thus providing clear
thermodynamic signatures of the topological
quantum phase transitions described above.

In summary, we investigated the effects of Zeeman fields
and ERD SOC on the superfluidity of two-dimensional
ultra-cold fermions from the BCS to the Bose regime.
We constructed the ground-state phase diagram
of Zeeman field versus spin-orbit coupling,
identified bulk topological (Lifshitz) phase transitions
between gapped and gapless superfluid.  We also
described the existence of a tri-critical point in the strong
spin-orbit coupling regime.
Lastly, we analyzed the excitation spectrum and the momentum distribution,
as well as thermodynamic quantities including the pressure, compressibility,
induced polarization, and spin-susceptibility, all of which provide signatures
of the phase transitions.

\acknowledgements{We thank Wei Zhang for discussions and
ARO (W911NF-09-1-0220) for support.}

\end{document}